\documentclass[]{article}
\usepackage{comment}
\usepackage{placeins}
\usepackage{graphicx}
\usepackage{subcaption}
\usepackage[dvipsnames]{xcolor}
\usepackage[normalem]{ulem}

\newcommand{\orcidID}[1]{\href{https://orcid.org/#1}{#1{\includegraphics[width=10pt]{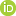}}}}

\usepackage{bbding}
\usepackage[pdfborder={0 0 0}]{hyperref}

\usepackage[]{authblk}
\usepackage{setspace}

\providecommand{\keywords}[1]
{
  \small	
  \textbf{\textit{Keywords---}} #1
}

\begin{document}

\title{A System for Automatic Rice Disease Detection from Rice Paddy Images Serviced via a Chatbot}

\author{
{Pitchayagan Temniranrat} [1] \orcidID{0000-0002-1951-7419} \\
{Kantip Kiratiratanapruk} [1] \orcidID{0000-0002-3083-0246} \\
{Apichon Kitvimonrat} [1] \orcidID{0000-0001-6161-5533} \\
{Wasin Sinthupinyo} [1] \orcidID{0000-0003-1804-3191} \\ 
{Sujin Patarapuwadol} [2] \orcidID{0000-0002-8337-4096}
}

\date{
    $^1$
 National Electronics and Computer Technology Center (NECTEC)
112 Phahonyothin Road, Khlong Nueng, 
Khlong Luang District, Pathumthani, Thailand 12120\\
 pitchayagan.temniranrat@nectec.or.th \\
 
$^2$
{Department of Plant pathology, Faculty of Agriculture at Kamphaeng Saen, Kasetsart University, Nakorn Pathom, Thailand\\
 agrsujp@ku.ac.th
\\[2ex]

}
}

\maketitle

\begin{abstract}

A LINE Bot System to diagnose rice diseases from actual paddy field images was developed and presented in this paper. It was easy-to-use and automatic system designed to help rice farmers improve the rice yield and quality. The targeted images were taken from the actual paddy environment without special sample preparation. We used a deep learning neural networks technique to detect rice diseases from the images. We developed an object detection model training and refinement process to improve the performance of our previous research on rice leave diseases detection. The process was based on analyzing the model’s predictive results and could be repeatedly used to improve the quality of the database in the next training of the model. The deployment model for our LINE Bot system was created from the selected best performance technique in our previous paper, YOLOv3, trained by refined training data set. The performance of the deployment model was measured on 5 target classes and found that the 
Average True Positive Point improved from 91.1\% in the previous paper to 95.6\% in this study.
Therefore, we used this deployment model for Rice Disease LINE Bot system. Our system worked automatically real-time to suggest primary diagnosis results to the users in the LINE group, which included rice farmers and rice disease specialists
. They could communicate freely via chat. In the real LINE Bot deployment, the model’s performance was measured by our own defined measurement Average True Positive Point and was found to be an average of 78.86\%. The system was fast and took only 2-3 seconds for detection process in our system server.

\end{abstract}

\keywords{Image Processing ,  Rice Disease, Deep Learning, Object Detection and Classification, Chatbot, LINE Bot and Neural Network Applications}

\section{Introduction}

Rice is an important economic crop and a food source for more than half of the world’s population. 36\% of staple food consumed over the world is rice and its consumption demand is predicted to be growing in the future \cite{OECDFAO}. The changing global climate, along with disease and pest outbreaks, poses a major and critical threat to rice production. The shortage of experts in plant diseases, tools and technology limits farmers who lack knowledge to solve problems in time. Therefore, it is necessary to have a simple and effective tool or technology to support and facilitate this critical issue for farmers.  

Moreover, with the huge growing popularity of camera-equipped mobile devices, the number of mobile applications for image analysis has been increased dramatically.
There are several examples of mobile applications related to agricultural and plant diseases, which can be downloaded through the Google Play Store. For example, Rice Doctor \cite{RiceDoctor} and riceXpert \cite{riceXpert} are mobile applications that offer both plant disease and pest diagnostics. They provide knowledge that displays images and text explanations. Also, keywords can be used to search information and frequently asked questions.

There are some applications that use artificial intelligence technology to diagnose plant diseases. One of them is Pestoz- \cite{pestoz}. Users can start to take a photo of diseases and send it to the diagnosis system. It will automatically process and return prediction results with solutions recommendation. An application called Plantix \cite{Rupavatharam2018} can support the identification of pests and plant diseases on over 30 plants. Agricultural officers and farmers can use it as a tool to advise crop prevention and proper management. However, there are still only a few applications for automatic diagnosis of plant diseases. Those applications were developed for each country’s plant species and disease types, with different purposes and needs.

However, there is a limitation to develop and maintain a standalone mobile application because the application’s behavior might change due to the user’s device OS version. In this paper, we reported on development and creation of a system based on LINE mobile application, a popular social communication application used in Asian countries. Developing system based on LINE application had no need to consider about the user’s device OS version, because LINE had service platform which did not depend on user’s device OS version. Our system used LINE Bot (automatic Chatbot worked with LINE account) for diagnosing rice diseases. 
Chatbot was an automatic robot that could perform as conversation partner for humans. There were various usages for Chatbot as described in Maroengsit et. al \cite{chatbot}. LINE Bot was used in various fields. Sudiatmika et. al \cite{balinese_LINE Bot} proposed a LINE Bot to translate Balinese language automatically. Boonsong \cite{netpiebot} used LINE Bot as communication tool between human users and IoT Platform. Sittakul et. al \cite{storagebot} access to on-site data storage by sending information and data through LINE Bot.

The information sent to LINE Bot would be passed to the disease diagnosis engine. The engine was supposed to detect disease objects in the paddy field images. Though there were various image processing research approaches on object detection as reviewed in Sharma et. al \cite{Sharma}, nowadays the research  had already moved to deep learning methods, which used convolution techniques and big size neural networks. 

The concept of training deep learning models is, to adjust the network's weights (also called the model's parameters) to minimize a loss function that measures miss prediction.
The model parameters are repetitively adjusted using back propagation to compute the gradient of the loss function that indicates the direction of change for the model's parameters. Usually, binary cross-entropy loss or its family is used for the binary classification problem and the catagorical cross-entropy loss or its family is used for the multi-class classification problem.
We deal with detection problem that not only classifies the object's class but also localizes it, therefore the regression loss is added to the cross-entropy loss.

This paper expanded on our previous published articles \cite{Kantip2020}, which was a comparative study of object detection techniques to assess the feasibility of their efficacy in diagnosing and predicting rice disease types. It was found that among 4 deep learning models: Faster R-CNN \cite{Ren2015FasterRT}, RetinaNet \cite{retinanet}, YOLOv3 \cite{Redmon2018YOLOv3AI} and Mask R-CNN \cite{He2017MaskR}, the best performance model in rice disease classification was YOLOv3.  

We proposed a rice plant disease diagnosis system from images taken in a real paddy field environment using mobile LINE application. Our system is practical and easy to use because the rice farmers can detect disease areas from images took from rice paddy field directly without any sample preparation or complicated tools. In addition, the diagnostic system has been designed and developed as a communication group where members are farmers and disease specialists, in order to exchange knowledge information, advice and solutions to problems between them. This will help diagnosis of the disease to be more efficient and more accurate. Our system can be operated real-time 24 hours everyday to give primary disease diagnosis results to the users.

\section{Methodology}

In the subsection \ref{sec:models}, we described a flow of training data preparation for rice diseases detection, object detection model training and refinement process. In the subsection \ref{sec:system}, we suggested an overview of our LINE Bot system followed by its details

\subsection{Rice Disease Detection Using Deep Learning Model}\label{sec:models}
\begin{figure}[hbt!]
\centering
\includegraphics[width=1.0\textwidth]{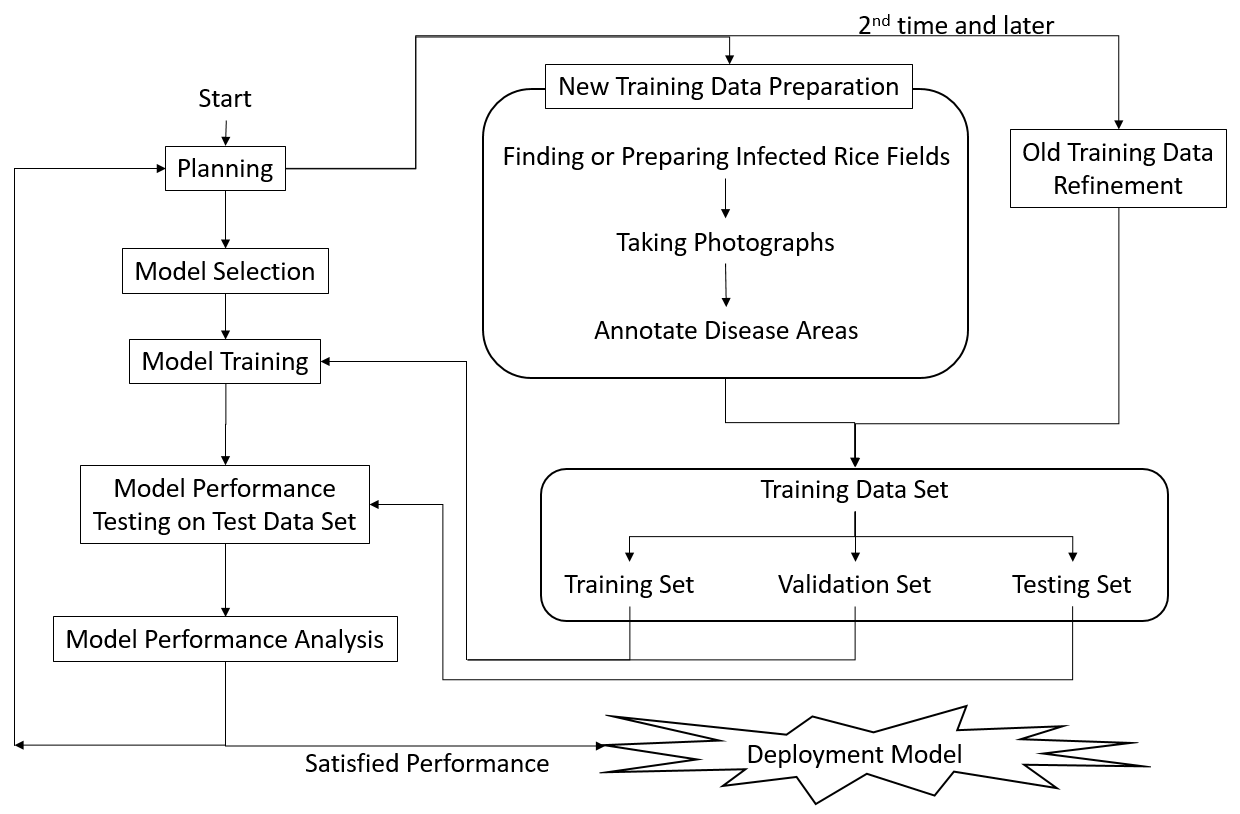}
\caption{Object Detection Model Training and Refinement Process} \label{fig:overviewmod}
\end{figure}

The overview of object detection model training and refinement process are shown in Figure \ref{fig:overviewmod}.

We started with planning how to prepare data and which model to be selected. We would call a deep learning model as a model later in this paper. We started our first planning in the previous research with a literature review. We then selected models, prepared training data set, made an experiment, and used the experimental results to help our planning repeatedly to improve the model performance.

Training data preparation process started with finding infected rice paddy fields or planting infected rice in a paddy field. This process was affected by several seasonal factors. 
Rice disease images were obtained by taking photographs from the fields.
Our diseases images were provided, examined and confirmed by panel of plant pathologist (later mentioned as rice specialists) from Kasetsart University and Thai Government's Rice Department, to avoid duplicates and ensure label quality.
The rice disease specialists then checked the quality of the images and categorized them based on the disease types appeared in each image.
We also confirmed with these specialists that even though the rice leaf diseases could be roughly divided by pathogen types (such as virus, bacteria and fungi) and the same type might cause similar symptoms, each pathogen actually caused different detailed physical symptoms that could be exactly visually classified.

The images would be annotated and the disease areas were labeled by rice specialists or researchers trained by the specialists. The obtained training data set would be divided in to 3 sets, training set, validation set and testing set.

The training set and the validation set would then be used to train the deep learning pre-trained model, and the testing set would be keep to evaluate the model's performance later.

In our previous research \cite{Kantip2020}, we selected 4 models including Faster R-CNN \cite{Ren2015FasterRT}, RetinaNet \cite{retinanet}, YOLOv3 \cite{Redmon2018YOLOv3AI} and Mask R-CNN \cite{He2017MaskR} to be tested and compared their performances as shown in Table \ref{tab:prev_mAP}.

In the object classification problem, the performance of a model could simply be measured as accuracy by comparing the ground-truth class and model's prediction class. However, in the object detection tasks, not only the class of the object was required to be predicted, but also the location of the object was.
Moreover, the trigger was set to determined which predicted result would be presented in the image, considering the prediction confident threshold. 
In order to measure the detection model's performance, precision and recall had been introduced. Precision measured how accurate the model can predict, and recall measured how well can the model trigger or detect the object in the image. There were 4 parameters used in precision and recall equation which were True Positive, False Positive, True Negative and False Negative, and when their object numbers were defined as TP, FP, TN and FN, precision was calculated by TP/(TP+FP) and recall was calculated by TP/(TP+FN).

Despite the fact that these measurements showed how well the model could classify and detect the object in the image, the bounding box's location was needed to be overlapped or matched with the ground-truth object location too. The simplest way to measure the bounding box localization performance was using Intersection over Union (IoU) method. IoU was calculated as the ratio of the overlap area and the union area of the predicted object area and the ground-truth object area.

PASCAL VOC project ~\cite{Everingham_2010} introduced measurement method to evaluate its provided image data set and annotations for comparison over different method based on Interpolated Average Precision (AP). This measurement calculated precision and recall of each object classes from all the image in the test data set where IoU is more than 0.5 . The result of this measurement gave mean Average Precision (mAP) for each object class calculated from an area under precision-recall graph, and average mAP represented the model's overall performance. 
We chose this Average mAP as performance measurement method, because it was widely used as measurement for object detection problem.

\begin{table}[hbt!]
\begin{center}
\caption{Mean Average precision (mAP) of the four models on detecting six rice diseases.}\label{tab:prev_mAP}

\begin{tabular}{l|cccc|}
\cline{2-5}
\multicolumn{1}{c|}{}                   & \multicolumn{4}{c|}{\textbf{Mean Average Precision (mAP)}}                                                                                       \\ \hline
\multicolumn{1}{|c|}{\textbf{Model}}    & \multicolumn{1}{c|}{\textbf{\begin{tabular}[c]{@{}c@{}}Faster\\R-CNN\end{tabular}}}                                                                                                                                & \multicolumn{1}{c|}{\textbf{RetinaNet}}                                                                                                                                & \multicolumn{1}{l|}{\textbf{YOLOv3}}  & {\textbf{\begin{tabular}[c]{@{}l@{}}Mask R-CNN\end{tabular}}}
\\ \hline
\multicolumn{1}{|c|}{\textbf{Backbone}} & \multicolumn{1}{l|}{\textbf{\begin{tabular}[c]{@{}c@{}}Inception\\V2\end{tabular}}} & \multicolumn{1}{l|}{\textbf{\begin{tabular}[c]{@{}c@{}}ResNet\\ 101\end{tabular}}} & 
\multicolumn{1}{l|}{\textbf{Darknet}} & \textbf{\begin{tabular}[c]{@{}l@{}}ResNet101\end{tabular}} \\ \hline
\multicolumn{1}{|l|}{\textbf{Blast}}    & 77.98                                                                                                                                                              & 41.87                                                                              & 86.46                                 & 84.95                                                         \\ \cline{1-1}
\multicolumn{1}{|l|}{\textbf{Blight}}   & 86.22                                                                                                                                                              & 27.84                                                                              & 90.33                                 & 88.83                                                         \\ \cline{1-1}
\multicolumn{1}{|l|}{\textbf{BSP}}      & 49.67                                                                                                                                                              & 36.97                                                                              & 56.88                                 & 49.37                                                         \\ \cline{1-1}
\multicolumn{1}{|l|}{\textbf{NBS}}      & 58.41                                                                                                                                                              & 55.46                                                                              & 83.72                                 & 81.07                                                         \\ \cline{1-1}
\multicolumn{1}{|l|}{\textbf{Streak}}   & 82.70                                                                                                                                                              & 30.21                                                                              & 96.33                                 & 89.55                                                         \\ \cline{1-1}
\multicolumn{1}{|l|}{\textbf{RRSV}}     & 70.75                                                                                                                                                              & 20.51                                                                              & 61.44                                 & 61.76                                                         \\ \hline
\multicolumn{1}{|l|}{\textbf{average mAP}}      & 70.96                                                                                                                                                              & 35.48                                                                              & 79.19                                 & 75.92                                                         \\ \hline
\end{tabular}

\end{center}
\end{table}

The experiment results from our previous research showed that YOLOv3 \cite{Redmon2018YOLOv3AI} gave the best performance to detect 6 classes of rice disease data, at mAP of 79.19\%. Therefore, in this paper, we selected YOLOv3 \cite{Redmon2018YOLOv3AI} as deep learning model architecture to be used in the developed LINE Bot system.

YOLOv3 \cite{Redmon2018YOLOv3AI} is a popular improvement version of YOLO. The concept of YOLO (You Only Look Once) \cite{Redmon2016YOLO} is to predict objects’ bounding boxes and class probabilities using only a single deep learning neural network in one evaluation. Therefore, YOLO can detect objects fast and is suitable for real-time applications. In YOLOv3, the bounding box prediction methods were improved and the convolutional neural network backbone’s size was enlarged.
YOLOv3 use binary cross-entropy loss function with logistic activation for object's class determination. In the object location bounding box localize part, logistic regression is used to predict objectness score and sum squared error is used as regression loss.

From our previous research results shown in Table. \ref{tab:prev_mAP}, we found that there were some classes that had remarkably low performance compared with other classes. After examination by the rice disease specialists, we found that there were 2 problems. The first problem was lack in symptom variety in the training data set. The second problem was improper data due to limited photographing environment. We returned to the planning process and improved the training data by adding data refinement process.

In our paper, the training data refinement was done by increasing data variety and removing improper data. The refinement details are shown in the Results and Discussion section. In the second time planning, we trained the selected model with a refined data set. We tested the model performance and analyzed test results with a consideration of comments from the rice disease specialists. When the specialists identified which model had satisfied performances, that model would be used as a deployment model. These processes could be repeated more than once and it was planned to be repeated again after the model was tested on the real LINE Bot system by real rice farmers in the actual environment.

\subsection{Design and Deployment of a LINE Bot System }\label{sec:system}

\begin{figure}[hbt!]
\centering
\includegraphics[width=1.0\textwidth]{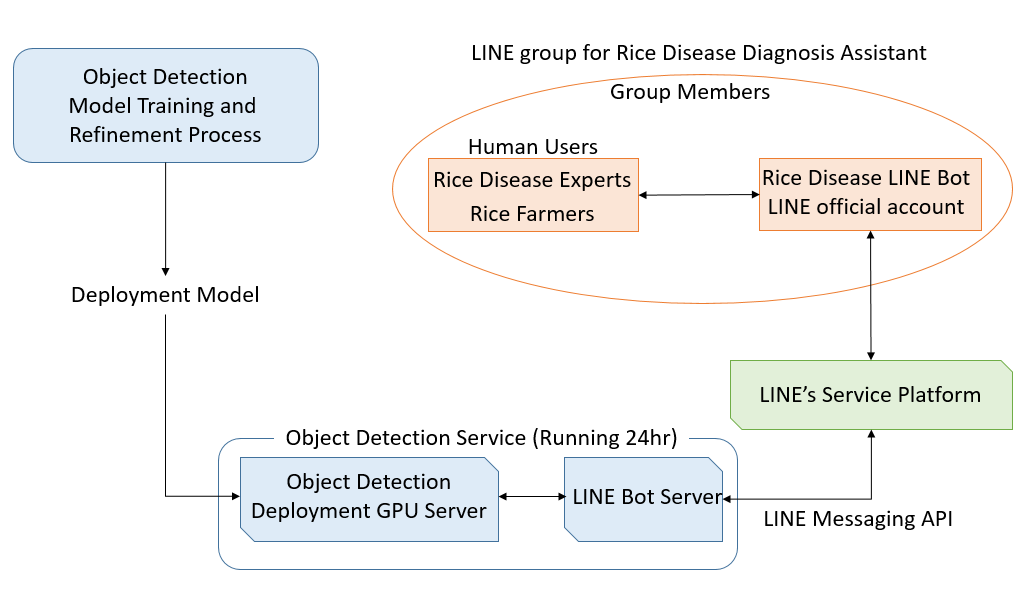}
\caption{LINE Bot System Overview} \label{fig:overviewsys}
\end{figure}

Our LINE Bot system was divided roughly into 2 parts as shown in Figure \ref{fig:overviewsys}. The first part was rice disease detection model training, which was detailed in the subsection \ref{sec:models}. The second part was rice disease detection LINE Bot deployment part.

We created LINE official account via https://developers.line.biz/en as a channel to communicate with target users in Thailand. The account was called “Bot account”. This account allowed our system to gain information from users and to send information back to the users through LINE’s service platform and LINE Messaging API. We created a social-network LINE group for rice disease diagnosis assistant called Rice Disease LINE Bot Group, which gave primary results of rice diseases diagnosis. There were disease detail documentations in the group’s album. The rice farmers and the rice disease specialists could communicate by chat freely. When our Bot account received an image sent by a group member, the image and its information (sender and time) would be sent through LINE’s service platform to our LINE Bot server. The LINE Bot server saved the image and the user’s ID information as a part of logs database.

The LINE Bot server would send the image to the object detection deployment GPU server, which contained a deep learning model loaded before hand and implemented object detection process. The deep learning model gave one or more locations of the object detected in the image and its class. We created one or more bounding box with its class and a prediction confidence as a label overlaid on the input image. The created image was sent back to the LINE Bot Server as a prediction result of an object detection. Additionally, results of the prediction class and its confidence described as text were also sent back as prediction results. The starting time and ending time of the detection process were saved in the logs database.

The LINE Bot Server sent prediction results in image and text as chat back through LINE Messaging API and LINE’s service platform to the Bot account. The LINE Bot server and the GPU server were running 24 hours for real-time services.
The users’ feedback and logs would be saved as data to refine our detection model in the future.

\section{Experimental Results}
We made an additional experiment to improve the model proposed in our previous paper [1] to be more accurate in actual environment. In the previous paper, we carried out experiments to train the model to classify 6 classes of rice diseases including blast, blight, brown spot (BSP), narrow brown spot (NBS), bacterial leaf streak (streak) and rice ragged stunt virus disease (RRSV). The sample numbers for each disease were shown in Table \ref{tab:previous_dataset}.

\begin{table}[hbt!]
\begin{center}
\caption{Training data set of six rice diseases in our previous research's experiment.}\label{tab:previous_dataset}
\resizebox{1.0\textwidth}{!}{

\begin{tabular}{lcllcll}
\hline
\multicolumn{1}{|l|}{\textbf{}} & \multicolumn{3}{c|}{\textbf{No. of label Box}} & \multicolumn{3}{c|}{\textbf{No. of Image Data}} \\ \hline
\multicolumn{1}{|l|}{\textbf{Class}} & \multicolumn{1}{l|}{\textbf{train}} & \multicolumn{1}{l|}{\textbf{validate}} & \multicolumn{1}{l|}{\textbf{Total}} & \multicolumn{1}{l|}{\textbf{train}} & \multicolumn{1}{l|}{\textbf{validate}} &
\multicolumn{1}{l|}{\textbf{Total}} \\ \hline
\multicolumn{1}{|l|}{\textbf{Blast}} & \multicolumn{1}{c|}{873} & \multicolumn{1}{c|}{217} & \multicolumn{1}{c|}{1,090} & \multicolumn{1}{c|}{805} & \multicolumn{1}{c|}{200} & 
\multicolumn{1}{c|}{1,005} \\
\multicolumn{1}{|l|}{\textbf{Blight}} & \multicolumn{1}{c|}{881} & \multicolumn{1}{c|}{214} & \multicolumn{1}{c|}{1,095} & \multicolumn{1}{c|}{866} & \multicolumn{1}{c|}{205} & 
\multicolumn{1}{c|}{1,071} \\
\multicolumn{1}{|l|}{\textbf{BSP}} & \multicolumn{1}{c|}{873} & \multicolumn{1}{c|}{216} & \multicolumn{1}{c|}{1,089} & \multicolumn{1}{c|}{513} & \multicolumn{1}{c|}{131} & 
\multicolumn{1}{c|}{644} \\
\multicolumn{1}{|l|}{\textbf{NBS}} & \multicolumn{1}{c|}{874} & \multicolumn{1}{c|}{214} & \multicolumn{1}{c|}{1,088} & \multicolumn{1}{c|}{822} & \multicolumn{1}{c|}{183} & 
\multicolumn{1}{c|}{1,005} \\
\multicolumn{1}{|l|}{\textbf{Streak}} & \multicolumn{1}{c|}{874} & \multicolumn{1}{c|}{215} & \multicolumn{1}{c|}{1,089} & \multicolumn{1}{c|}{829} & \multicolumn{1}{c|}{204} & 
\multicolumn{1}{c|}{1,033} \\
\multicolumn{1}{|l|}{\textbf{RRSV}} & \multicolumn{1}{c|}{873} & \multicolumn{1}{c|}{214} & \multicolumn{1}{c|}{1,087} & \multicolumn{1}{c|}{682} & \multicolumn{1}{c|}{170} & 
\multicolumn{1}{c|}{852} \\ \hline
\multicolumn{1}{|l|}{\textbf{Total}} & \multicolumn{1}{c|}{5,248} & \multicolumn{1}{c|}{1,290} & \multicolumn{1}{c|}{6,538} & \multicolumn{1}{c|}{4,517} & \multicolumn{1}{c|}{1,093} & 
\multicolumn{1}{c|}{5,610} \\ \hline
\multicolumn{1}{|l|}{\textbf{Total (\%)}} & \multicolumn{1}{c|}{80.27} & \multicolumn{1}{c|}{19.73} & 
\multicolumn{1}{c|}{100} & \multicolumn{1}{c|}{80.52} & \multicolumn{1}{c|}{19.48} & 
\multicolumn{1}{c|}{100} \\ \hline

\end{tabular}
}
\end{center}
\end{table}

After the previous result analysis, it was found that the RRSV class performed poorly because the images were taken in a greenhouse, not in a paddy farm where other class images were taken. The greenhouse environment was not our supposed target environment. Therefore, we removed the RRSV class from this experiment to improve the accuracy of the model. In addition, BSP class also performed lower than the other classes due to seasonal restriction causing limited numbers of symptom variety. We analyzed the training data set and made an assumption that all BSP images were taken when the disease was in the early stage symptom and the symptom was not visualized clearly. In the new training data set, more varieties of symptom stages were included.

We increased the size of training data set to about double of the numbers in order to improve classification performance and trained the YOLOv3 model by this refined training data set as shown in Table. \ref{tab:samplenumber}. The comparison of numbers of Training Data of 5 disease classes in the previous research and in this paper were shown in Table. \ref{tab:comparenumber}

\begin{table}[hbt!]
\begin{center}
\caption{Training data set of five rice diseases in this paper's experiment.}
\label{tab:samplenumber}
\resizebox{1.0\textwidth}{!}{
\begin{tabular}{lcllcll}
\hline
\multicolumn{1}{|l|}{\textbf{}} & \multicolumn{3}{c|}{\textbf{No. of label Box}} & \multicolumn{3}{c|}{\textbf{No. of Image Data}} \\ \hline
\multicolumn{1}{|l|}{\textbf{Class}} & \multicolumn{1}{l|}{\textbf{train}} & \multicolumn{1}{l|}{\textbf{validate}} & \multicolumn{1}{l|}{\textbf{Total}} & \multicolumn{1}{l|}{\textbf{train}} & \multicolumn{1}{l|}{\textbf{validate}} & \multicolumn{1}{l|}{\textbf{Total}} \\ \hline
\multicolumn{1}{|l|}{\textbf{Blast}} & \multicolumn{1}{c|}{1,622} & \multicolumn{1}{c|}{351} & \multicolumn{1}{c|}{1,973} & \multicolumn{1}{c|}{1,490} & \multicolumn{1}{c|}{331} & \multicolumn{1}{c|}{1,821} \\ 
\multicolumn{1}{|l|}{\textbf{Blight}} & \multicolumn{1}{c|}{1,641} & \multicolumn{1}{c|}{351} & \multicolumn{1}{c|}{1,992} & \multicolumn{1}{c|}{1,583} & \multicolumn{1}{c|}{325} & \multicolumn{1}{c|}{1,908} \\ 
\multicolumn{1}{|l|}{\textbf{BSP}} & \multicolumn{1}{c|}{1,606} & \multicolumn{1}{c|}{352} & \multicolumn{1}{c|}{1,958} & \multicolumn{1}{c|}{1,213} & \multicolumn{1}{c|}{253} & \multicolumn{1}{c|}{1,466} \\ 
\multicolumn{1}{|l|}{\textbf{NBS}} & \multicolumn{1}{c|}{1,642} & \multicolumn{1}{c|}{351} & \multicolumn{1}{c|}{1,993} & \multicolumn{1}{c|}{1,433} & \multicolumn{1}{c|}{318} & \multicolumn{1}{c|}{1,751} \\ 
\multicolumn{1}{|l|}{\textbf{Streak}} & \multicolumn{1}{c|}{1,604} & \multicolumn{1}{c|}{351} & \multicolumn{1}{c|}{1,955} & \multicolumn{1}{c|}{1,494} & \multicolumn{1}{c|}{327} & \multicolumn{1}{c|}{1,821} \\ \hline
\multicolumn{1}{|l|}{\textbf{Total}} & \multicolumn{1}{c|}{8,115} & \multicolumn{1}{c|}{1,756} & \multicolumn{1}{c|}{9,871} & \multicolumn{1}{c|}{7,213} & \multicolumn{1}{c|}{1,554} & \multicolumn{1}{c|}{8,767} \\ \hline
\multicolumn{1}{|l|}{\textbf{Total (\%)}} & \multicolumn{1}{c|}{82.21} & \multicolumn{1}{c|}{17.79} & \multicolumn{1}{c|}{100} & \multicolumn{1}{c|}{82.27} & \multicolumn{1}{c|}{17.73} & \multicolumn{1}{c|}{100} \\ \hline

\end{tabular}
}
\end{center}
\end{table}

\begin{table}[hbt!]
    \centering
    \caption{Comparison of numbers of training data of 5 disease classes in the previous research and in this paper}
    \label{tab:comparenumber}
    \begin{tabular}{|l|c|c|c|c|}
        \hline
        \multicolumn{1}{|c|}{} & \multicolumn{2}{c|}{\textbf{No. of label Box}} & \multicolumn{2}{c|}{\textbf{No. of Image Data}} \\ 
        \hline
        \multicolumn{1}{|c|} {\textbf{Class}} & \multicolumn{1}{|c|} {\textbf{Prev. Model}} & \multicolumn{1}{|c|} {\textbf{This Paper}} & \multicolumn{1}{|c|} {\textbf{Prev. Model}} & \multicolumn{1}{|c|} {\textbf{This Paper}} \\
        \hline
        \textbf{Blast} & 1,090 & 1,973 & 1,005 & 1,821 \\
        \textbf{Blight} & 1,095 & 1,992 & 1,071 & 1,908 \\
        \textbf{BSP} & 1,089 & 1,958 & 644 & 1,466 \\
        \textbf{NBS} & 1,088 & 1,993 & 1,005 & 1,751 \\
        \textbf{Streak} & 1,089 & 1,955 & 1,033 & 1,821 \\ \hline
        \textbf{Sum} & 5,451 & 9,871 & 4,758 & 8,767 \\ \hline
    \end{tabular}

\end{table}

We trained the model using Machine with the new data set on NVIDIA DGX1, Dual 20-Core Intel Xeon E5-2698 v4 2.2 GHz CPU, 512 GB 2,133 MHz, 8X NVIDIA Tesla V100 GPU. The architecture of the model was the same as the previous model. 

We found that the location of the object detection was not the main requirement in actual usage from examining the users’ behavior. This meant that mAP did not provide the performance that the users would like to know. So, we defined a new accuracy measurement, called Average True Positive Point that indicated whether the disease shown in the images was classified correctly. If the picture had only one disease class and the prediction class was the same as the ground truth class, it would gain 1 point. If 
there was no prediction or
the prediction class differed from the ground truth class, it would gain 0 point. If the picture had more than one class and all prediction classes were the same as the ground truth classes, it would gain 1 point, and if 
there was no prediction or
all prediction classes differed from the ground truth class, it would gain 0 point. If the number of prediction class appeared in the ground truth was n and the number of prediction class not appeared in ground truth was m, it would gain n/(n+m) point. 
And in the case that both n and m is zero, it would get 0 point.
The Average True Positive Point was defined as the summation of all image points in each class divided by each class number of the images.

The model performance measured on the test data set in which no image is the member of the training data set by 
Average True Positive Point for each class were represented in Table. 
\ref{tab:compareTPpoint}. Note that the Average True Positive Point of the BSP was remarkably improved from 83.5\% to 90.0\% and NBS also remarkably improved from 83.0\% to 95.5\%. Most importantly, the Average True Positive Point of all classes were improved and the average Average True Positive Point of all five classes was improve from 91.1\% to 95.6\%.

\begin{table}[hbt!]
    \centering

    \caption{Disease prediction performance comparison between the previous research's model and the model refined in this paper in Average True Positive Point (\%)}
    \label{tab:compareTPpoint}
\begin{tabular}{|l|c|c|c|}
\hline
\textbf{Class} & \textbf{The No. of test}      & \multicolumn{2}{l|}{\textbf{Average True Positive Point (\%)}} \\ \cline{3-4} 
                       & \textbf{Rice Disease Images} & \textbf{Previous Research}             & \textbf{This Paper}            \\ \hline
\textbf{Blast}                  & 100                 & 94.5                          & 95.5                  \\
\textbf{Blight}                 & 100                 & 99.0                            & 100                   \\
\textbf{BSP}                    & 100                 & 83.5                          & 90.0                    \\
\textbf{NBS}                    & 100                 & 83.0                            & 95.5                  \\
\textbf{Streak}                 & 100                 & 95.5                          & 97.0                    \\ \hline
\textbf{Total}                  & 500                 & 91.1                          & 95.6                  \\ \hline
\end{tabular}
\end{table}

We then used this model in our newly developed LINE Bot system as a deployment model. The GPU server machine for object-detection specification was Intel(R) Xeon(R) Gold 6144 CPU @ 3.50GHz 32 core RAM 125 GB, Tesla P100-PCIe 16 GB GPU. The LINE Bot server specification was Intel(R) Xeon(R) Gold 5115 CPU @ 2.40GHz 4 core RAM 3.9 GB. Our LINE group had about 470 members, including rice disease specialists and rice farmers. Our LINE bot replied to all images submitted by this group. The pictures sent to the system were taken from various actual environment and places, and by varieties of cameras.

We measured the model performance based on the conversation log in the LINE group. The rice disease specialists in the LINE group not only gave suggestions but also helped verifying the model performance and pointed out false predictions. The detection results, which the rice specialists pointed out to be wrong, were determined to be fault detection.

The chat logs between 22 November 2019 and 15 March 2020 were used to measure preliminary results. The usage manual of our system clearly notified the users that the input image should be a focused image of rice leaves taken from a rice paddy field and our system supported only 5 diseases as prescribed. There were quite a number of images that were out of these specifications among 1,574 images sent to our system. Approximately 55\% of the images were target diseases images (868 images) and 45\% were not. The non-target images included 294 images that were blurred or not taken from a rice field, and 412 images had anomaly caused by others diseases or chemical factors or presence of some pests. In the experiment, we only focused on the target images.

Disease prediction performance in real LINE Bot usage environment in Average True Positive Point (\%) is shown as in Table. \ref{tab:bot_exp}.

\begin{table}[hbt!]
    \centering
    \caption{Disease prediction performance in real LINE Bot usage environment in Average True Positive Point (\%)}
    \label{tab:bot_exp}
    \begin{tabular}{|l|c|c|c|}
        \hline
        Class	 &   The No. of  &	True Positive Point &	Average True \\
        & Rice Disease Images & & Positive Point (\%)  \\
        \hline
        Blast &	182 &	146&	80.22 \\
        Blight &	191 &	162 &	84.82 \\
        BSP &	346 &	235 &	67.92 \\
        NBS &	122 &	116.5 &	95.49 \\
        Streak &	27 &	25 &	92.59 \\ \hline
        Total &	868 &	684.5 &	78.86 \\
        \hline
    \end{tabular}
\end{table}

\begin{figure}[hbt!]
\centering
\includegraphics[width=0.7\textwidth]{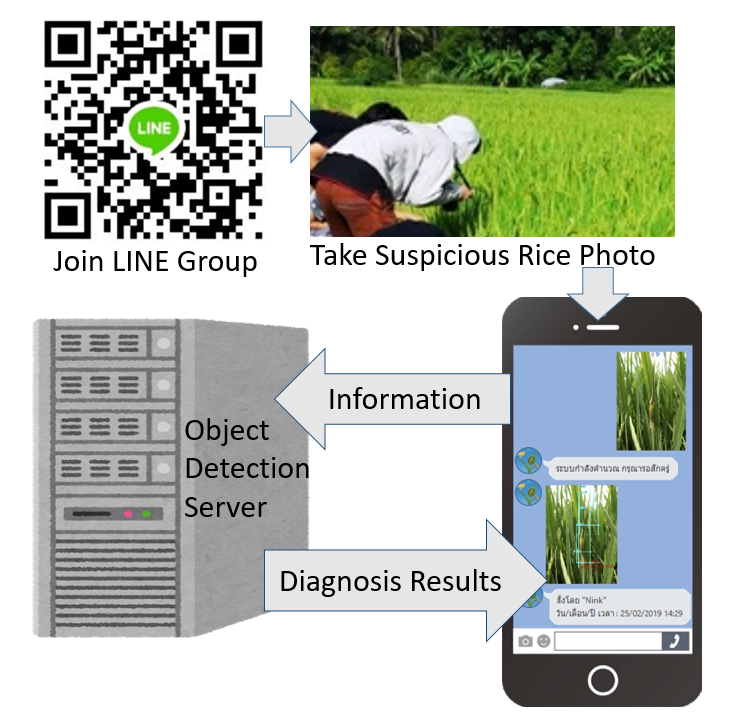}
\caption{LINE Bot System Usage Sample}
\label{fig:UsageSample}
\end{figure}

\begin{figure}[hbt!]
\begin{subfigure}{0.45\textwidth}
\centering
\includegraphics[height=6cm]{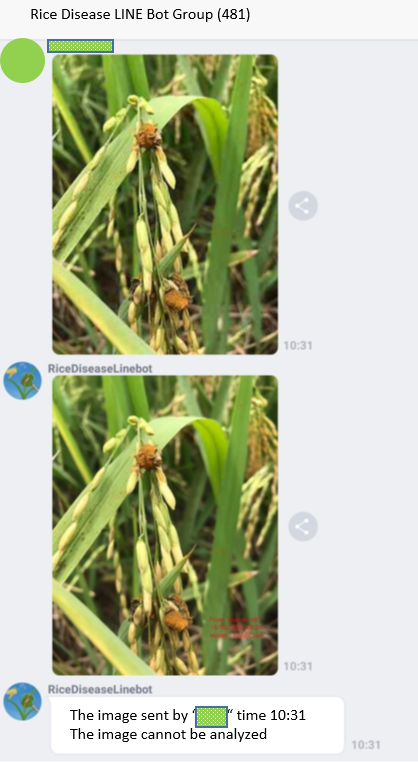} 
\caption{Bot correctly not detect \\ non-targeted object}
\label{fig:Bot_ex_nontarget}
\end{subfigure}
\begin{subfigure}{0.45\textwidth}
\centering
\includegraphics[height=6cm]{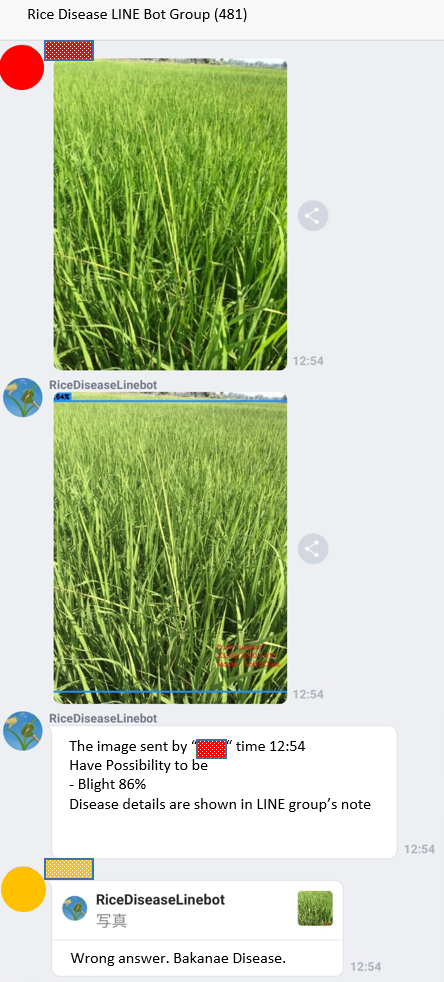}
\caption{Wrong Bot Answer corrected \\ by rice expert}
\label{fig:Bot_ex_wrong}
\end{subfigure}

\begin{subfigure}{1.0\textwidth}
\centering
\includegraphics[width=0.45\linewidth]{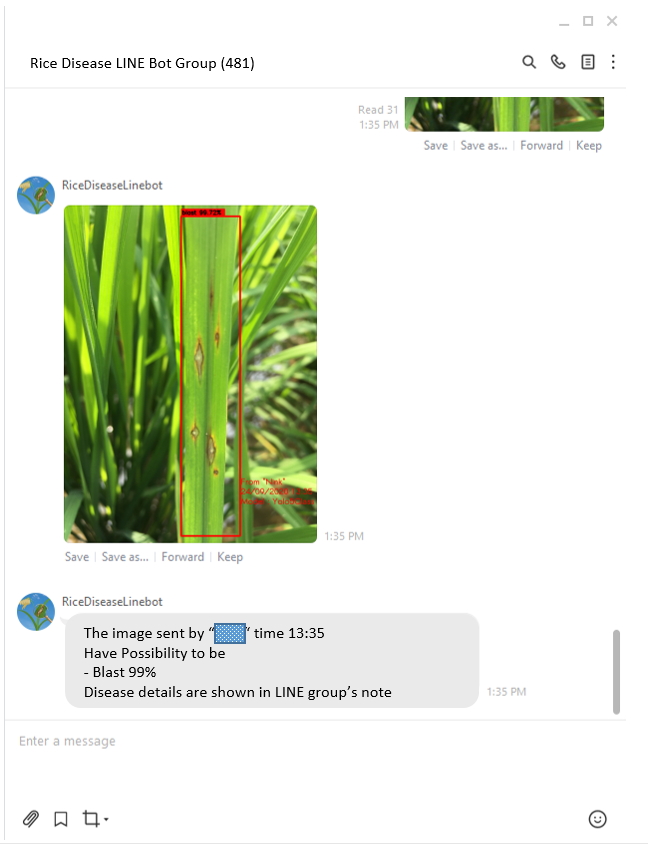} 
\caption{Bot Answer Correctly}
\label{fig:Bot_ex_right}
\end{subfigure}

\caption{LINE Bot System Real Usage Chat Dialog Translated into English} \label{fig:LINE Bot_ex}
\end{figure}

It took approximately 2-3 seconds from the time an image arrived at our LINE Bot server until our LINE Bot server started sending disease detection prediction result image back to the LINE Service Platform. If the user’s network environment did not have any problem, it took about 5 seconds from the time the image was sent to the LINE official account until the time the user received prediction result back. The sample image of the actual usage on the LINE Bot was shown in Figure. \ref{fig:UsageSample} and the chat dialog details translated into English was shown in Figure. \ref{fig:LINE Bot_ex}.

We could analyze and divide Bot responses and conversation flow in the LINE group into 3 types, 1. Did not answer to the image that do not meet the system requirement (Figure \ref{fig:Bot_ex_nontarget}), 2. Predicted the wrong answer and the rice specialists gave the correction via chat (Figure \ref{fig:Bot_ex_wrong}) and 3. Detected disease area and answered correctly (Figure \ref{fig:Bot_ex_right}) .
The LINE Bot's performance analysis process was made by using rice disease specialists' model's prediction results verification.

Results from a questionnaire to determine satisfaction of our system prediction performance, sent to six rice specialists indicated that from 4 steps satisfaction level questionnaire, 0 voted as not satisfied, 1 voted as fair, 5 voted as moderate, and 0 voted as good. 

For our system to be more practical in the future, we are planning to increase target diseases and each disease’s symptom variety. We are also planning to develop a server maintenance system that can easily handle more users and LINE groups without specific technical knowledge about database or LINE Bot. Furthermore, we are planning to create a self refinement detection model using the specialists’ feedback about false detection, in order to improve the system’s disease detection performance.

\section{Conclusion}
In this paper, we developed a LINE Bot System used deep learning techniques to detect rice diseases from images taken in a paddy field. We trained the model architecture chosen as an extension of our previous research. Training data refinement was made systematically. The training data images were increased and the class that had improper environment images was removed. The performance of the deployment model improved from 
91.1\% to 95.6\%. The model was trained and tested by the new data set measured by Average True Positive Point of 5 targeted classes.
In the real LINE Bot deployment, the model performance measured by Average True Positive Point was 78.86\%. It took only 2-3 seconds for detection process in our system server. 

In the future works, we are planning to enable individual chat and add location tagging function, to use disease spreading information more effectively. 
We are obtaining training data with more variation and planning to refine the system's disease detection model to be able to analyze more disease type. Our aim is to be able to analyze most diseases frequently found in Thailand.
Now that our system users amount is continuously increasing, the system might confront both expected and unexpected problems. Therefore, we are planning to create system maintenance assistant application that will help system administrator manage the LINE Bot and chat community easily. 
We are designing to extract the rice specialists' chat dialog details automatically, and use them to obtain more data to improve the model's performance.

\section{Acknowledgement}
This study was a part of ”Mobile application for rice disease diagnosis using image analysis and artificial intelligence” project, supported by grants from Innovation for Sustainable Agriculture (ISA), National Science and Technology Development Agency, Thailand [grant number P18-51456]. We would like to thank the team from Kamphaeng Saen, Kasetsart University, who supported our data preparation and provided plant disease diagnosis knowledge. We are also grateful to Prof. S. Seraphin (Professional Authorship Center, NSTDA) for fruitful discussion on preparing this manuscript


\begin{thebibliography}{16}

\bibitem{OECDFAO}
OECD/FAO (2020), OECD-FAO Agricultural Outlook 2020-2029, OECD Publishing, Paris/FAO, Rome, DOI:https://doi.org/10.1787/1112c23b-en.

\bibitem{RiceDoctor}
LucidMobile. (2017). Rice Doctor (Version 1.0.10) [Mobile app]. Retrieved from https://play.google.com/

\bibitem{riceXpert}
National Rice Research Institute. (2020). riceXpert (Version 3.7) [Mobile app]. Retrieved from https://play.google.com/

\bibitem{pestoz}
AgroConnectIndia. (2017). Pestoz- Identify Plant diseases (Version 1.1.58) Retrieved from https://play.google.com/

\bibitem{Rupavatharam2018}
Rupavatharam S, Kennepohl A, Kummer B, Dhulipala R (2018) An android mobile application to detect rice diseases in field and support smallholder farmers, International Rice Congress, Oct 2018.

\bibitem{chatbot}
Maroengsit W, Piyakulpinyo T, Phonyiam K, Pongnumkul S, Chaovalit P, and Theeramunkong T. (2019). A Survey on Evaluation Methods for Chatbots, In Proceedings of the 2019 7th International Conference on Information and Education Technology (ICIET 2019), Association for Computing Machinery, New York, NY, USA, 111–119. DOI:https://doi.org/10.1145/3323771.3323824

\bibitem{balinese_LINE Bot}
Sudiatmika IPGA, Putra IMAW, Dewi KHS, Aryawan IKB (2019) Line Bot Implementation for Automation Balinese Language Dictionary, 2019 1st International Conference on Cybernetics and Intelligent System (ICORIS), Denpasar, Bali, Indonesia, 2019, pp. 227-232, DOI:10.1109/ICORIS.2019.8874907.

\bibitem{netpiebot}
Boonsong W (2019) Smart Intruder Notifying System Using NETPIE through Line Bot Based on Internet of Things Platform, IEEE 5th International Conference on Computer and Communications (ICCC), Chengdu, China, 2019, pp. 2208-2211, DOI:10.1109/ICCC47050.2019.9064030.

\bibitem{storagebot}
Sittakul V, Khotwongsa W, Poolthep Y and Pasakawee S (2019) On-site Data Storage via Website or LineBOT, IEEE Asia Pacific Conference on Circuits and Systems (APCCAS), Bangkok, Thailand, 2019, pp. 409-412, DOI:10.1109/APCCAS47518.2019.8953086

\bibitem{Sharma}
Sharma KU, Thakur NV (2017). A review and an approach for object detection in images, International Journal of Computational Vision and Robotics. 7. 196. DOI:10.1504/IJCVR.2017.081234.

\bibitem{Kantip2020}
Kiratiratanapruk K, Temniranrat P, Kitvimonrat A, Sinthupinyo W, Patarapuwadol S (2020) Using Deep Learning Techniques to Detect Rice Diseases from Images of Rice Fields, In: Trends in Artificial Intelligence Theory and Applications. Artificial Intelligence Practices. IEA/AIE 2020. Lecture Notes in Computer Science, vol 12144. Springer, Cham. DOI:https://doi.org/10.1007/978-3-030-55789-8\_20

\bibitem{Ren2015FasterRT}
Ren S, He K, Girshick RB, Sun J (2015) Faster R-CNN: Towards Real-Time Object Detection with Region Proposal Networks, IEEE Transactions on Pattern Analysis and Machine Intelligence, vol.39, pp.1137--1149, 2015

\bibitem{retinanet}
Lin TY, Goyal P, Girshick RB, He K, Doll{\'{a}}r (2017) P Focal Loss for Dense Object Detection, ArXiv, 2017

\bibitem{He2017MaskR}
He K, Gkioxari G, Doll{\'a}r P, Girshick RB (2017) Mask R-CNN, IEEE International Conference on Computer Vision (ICCV), pp.2980--2988, 2017 

\bibitem{Redmon2018YOLOv3AI}
Redmon J, Farhadi A (2018) YOLOv3: An Incremental Improvement, ArXiv, 2018

\bibitem{Redmon2016YOLO}
Redmon J, Divvala S, Girshick R, Farhadi A (2016) You Only Look Once: Unified, Real-Time Object Detection. 779-788. DOI:10.1109/CVPR.2016.91. 

\bibitem{Everingham_2010}
Everingham M, Gool L, Williams CK, Winn J, Zisserman A (2010) The Pascal Visual Object Classes (VOC) Challenge, International Journal of Computer Vision, Vol.88, pp.303–-338, 2010, DOI:https://doi.org/10.1007/s11263-009-0275-4

\end{thebibliography}
\end{document}